\documentclass{article}
\usepackage{a4wide}
\usepackage{graphicx}
\usepackage[utf8]{inputenc}
\usepackage{amsmath}
\usepackage{amssymb}
\usepackage{mathtools}
\usepackage{tabularx}
\usepackage{multirow}
\usepackage{rotating}
\usepackage[style=chem-acs,backend=bibtex]{biblatex}
\bibliography{mybib.bib}

\newcommand{\zr}{\begin{equation}}
\newcommand{\kr}[1]{\label{#1} \end{equation}}

\title{Simulations of coherent nonlinear optical response of molecular vibronic dimers}
\author{Václav Perlík, František Šanda \\
	Faculty of Mathematics and Physics, Charles University in Prague  \\
vaclav.perlik@gmail.com
	}

\date{\today}

\begin{document}

\maketitle

\begin{abstract}
We have implemented vibronic dynamics for simulations of the third order coherent response of electronic dimers.
In the present communication  we provide the full and detailed description of the dynamical model,
recently used for simulations of chlorophyll-carotenoid dyads, terylene dimers, or hypericin.
We allow for explicit vibronic level structure, by including selected vibrational modes into a "system".
Bath dynamics include the Landau-Teller vibrational relaxation, electronic dephasing, and nonlinear vibronic (to bath) coupling.
Simulations combine effects of transport and dephasing between vibronic levels. Transport is described by master equation  within secular approximation, phase is accumulated in cumulants and its calculation follows the transport pathways during waiting time period.
\end{abstract}

\section{Introduction}
Two dimensional optical spectra \cite{mukamel2000multidimensional,jonas2003two,cho2009two} of molecular aggregates shows almost invariantly important role of vibrations on relaxation and electronic transport \cite{perlik2015vibronic} within organic dyes, light harvesters and similar molecular systems of intensive interest. Interplay of vibrational and electronic dynamics is a computational challenge which has been approached on the various level of theory \cite{abramavicius2009coherent}.

Description of nonlinear response \cite{mukamel1999principles} of transporting multichromophoric electronic systems was pioneered  by Zhang, Meier, Chernyak and Mukamel \cite{zhang1998exciton}. Algorithm is based on partitioning bath induced fluctuations of excitons into "diagonal part", that is, frequency (eigenfrequency) fluctuations and off-diagonal, that is coupling fluctuations, the former accounted for by cumulants and responsible for lineshapes, the latter using some kind of reduced dynamics  (master equations) and responsible for transport. This strategy after some refinements became standard, the choice of transport dynamics, bath models  and further details of simulations, however differs among authors.  More recent developments, and other strategies (important, but not relevant for the present communication) for simulating 2D spectra can be found in reviews  \cite{abramavicius2009coherent,mukamel2004many,falvo2013quasi}.

Calculations along such a strategy work quite well, when the vibrational modulation are broad in frequencies and the spectra are thus without clear signatures of underdamped mode such as vibronic progression in the spectra.    A better representation of underdamped mode requires to make them explicit, i.e. exclude important vibrational modes from bath and include them formally into the "system" together with electronic degrees of freedom and  calculate transport rates between the mixed electronic-vibrational levels. While direct implementation  is costly and is thus of limited use for extended aggregates, it is accessible for small aggregates.  After redefining the system, however, the rest of simulation strategy instituted by Ref. \cite{zhang1998exciton}  can remain largely  untouched. Various variant of such a strategy has been simultaneously followed recently by several authors \cite{butkus2014vibronic,butkus2013distinctive,polyutov2012exciton,schroter2015exciton}.

In the present communication we report a dynamical model of vibronic dynamics affordable for small (dimers, trimers) molecular aggregates with resolved vibrational structure
which we have implemented recently to inquire phase relation in vibronic systems \cite{perlik2014distinguishing}, and further used it for simulation of 2D spectra  of carotenoid-chlorophyll dyad \cite{perlik2015vibronic}, and transient grating on hypericin \cite{lincoln2015A}. Few others applications (perylene dimers) are yet in preparation \cite{perlik2016perylene}. While the vibronic structures and system-bath dynamics were reported in the respective publications, there remains couple of interesting details of our implementation  to be clarified to public. Last but not least, we have to report the full potential of our code yet.

So far, we have implemented and applied the vibronic dynamics only for molecular dimers. However, we are intending to apply vibronic dynamics to some trimeric systems.
In addition, there would be no significant simplification of this manuscript if we limited the introduced formalism just to dimer. We thus  define the vibronic dynamics for a general aggregate in sections \ref{vibronicsystem} to \ref{liouvillespacepathways}, keeping in mind that our implementation was actually tested only for dimers so far.

The paper is structured as follows. In section \ref{vibronicsystem} we introduce model of vibronic aggregate.  System is separated from bath and diagonalized.
In section \ref{interactionwithbath} we introduce bath induced fluctuations. Dynamics include Landau-Teller vibrational relaxation, electronic dephasing and nonlinear bath-to vibration Hamiltonian. Diagonal eigenfrequency fluctuations are distinguished from off-diagonal eigenstate fluctuations.
In section \ref{eigenstatefluctuations} we specify effect of eigenstate fluctuations and describe transport dynamics using secular time-convolutionless master equation.
In section \ref{eigenfrequencyfluctuations} we take care of eigenfrequency fluctuations and define their correlation and lineshape functions.
In section \ref{liouvillespacepathways} we calculate linear and the third order optical response. We show Feynman diagrams, and
calculate second cumulants for eigenfrequency fluctuations along the diagrams. In our implementation we follow to some extent the reorganization of bath modes during transport. Section \ref{rotationalaveraging} discuss tensorial character of nonlinear response and rotational averaging of a response from randomly oriented chromophores in an isotropic sample.
In section \ref{spectracalculations} we linked response functions to physical observable of interest: absorption, fluorescence, 2D electronic spectra, frequency resolved transient grating and pump probe are calculated.
In section \ref{applicatons} we connect the previous sections with our recent simulations.
In section \ref{conclusion} we conclude.

\section{Vibronic System} \label{vibronicsystem}

We simulate nonlinear optical response of molecular aggregates with significant vibronic structure. Electronic structure of each molecule ($i=1,2,...,N$) of the aggregate is modelled by a two level chromophore (with ground $g_i$ and excited $e_i$ levels, and transition (gap) frequency $\epsilon_i$). For the third order response only a limited part of the composed electronic Hilbert space of aggregate is relevant. The relevant part consists of a ground state $|\Pi_i g_i\rangle$, one-exciton states where a single chromophore is excited $|e_k \Pi_{i\neq k}g_i\rangle = \hat{A}^{\dag}_k |\Pi_i g_i\rangle$ and a doubly excited states $|e_k e_l \Pi_{i\neq k,l} g_i\rangle  = \hat{A}^{\dag}_k \hat{A}^{\dag}_l |\Pi_i g_i\rangle$.
 The molecules are further resonantly coupled; in a standard approximation the number of exciton is conserved and excitation of $i$-th chromophore is associated with deexcitation of some $j$-th chromophore. The Frenkel exciton (electronic) Hamiltonian $\hat{H}_e$  is thus
 \begin{align}
 \label{FE}
 \hat{H}_e = \hbar \sum_i \epsilon_i \hat{A}_i^{\dag} \hat{A}_i + \hbar\sum_{i\neq j} J_{ij} \hat{A}^{\dag}_i \hat{A}_j
 \end{align}
 While the determination of transition frequencies does not pose significant problems, strategies for estimate of intermolecular coupling  $\hbar J_{ij}$ differs. It is approximated by dipole-dipole force, calculated using Quantum chemistry libraries or fitted from the spectra.

Quantum dynamics (defined by Eq. (\ref{FE})) within the three electronic manifolds forms standard approaches for calculations of third order response.
 In the present work we have included certain vibrations as part of the system. Vibrations are assumed local. Each vibration is attached to some chromophore, $q_{i,z}$ is the coordinate of $z$-th mode on $i$-th chromophore. We assumed electronic potential surface to be harmonic with respect to nuclear coordinates $q_{i,z}$, thus $U_{i,z}(q_{i,z}) = \frac{1}{2} m_{i,z} \omega_{i,z}^2 q_{i,z}^2$ is the ground state's potential and $\tilde{U}_{i,z}(q_{i,z}) = \hbar \epsilon_i + \frac{1}{2} m_{i,z} \omega_{i,z}^2(q_{i,z} - d_{i,z})^2$ is the excited states' potential surface, where $\omega_{i,z}$ is the vibrational frequency, $m_{i,z}$ is the mass and $d_{i,z}$ is displacement.
Vibrational Hamiltonian $\hat{H}_v$  reads
\begin{align}\hat{H}_v = \sum_{i,z} \hbar \omega_{i,z}(\hat{V}_{i,z}^{\dag} \hat{V}_{i,z} + 1/2) + \frac{1}{2}m_{i,z}\omega^2_{i,z} d_{i,z} \left( -2\sqrt{\frac{\hbar}{2m_{i,z}\omega_{i,z}}}(\hat{V}_{i,z}^{\dag} + \hat{V}_{i,z}) + d_{i,z}\right)  \hat{A}^{\dag}_i \hat{A}_i
\end{align}
where $\hat{V}^{\dag}_{i,z}$ $(\hat{V}_{i,z})$ are creation (annihilation) operators of $q_{i,z}$ harmonic vibrational mode.

Vibrons are complex electronic-vibrational excitations. To describe them we have to define system as composed from electronic and vibrational degrees of freedom. We thus works in composed Hilbert space and the system molecular Hamiltonian $\hat{H}_S$  is
\begin{align}
\label{vibron_ham}
\hat{H}_S=\hat{H}_e+\hat{H}_v
\end{align}

We next introduce convenient basis. We start with basis of vibrational states over site. It consist of  ground  state electronic wavefunctions $g_i$, with a  well-known wave functions of harmonic oscillators
\zr |g_i \Pi_{z} n_{i,z}\rangle= |g_i\rangle \otimes \prod_{z}\frac{1}{\sqrt{n_{i,z}!}} \left(\hat{V}_{i,z}^{\dag}\right)^{n_{i,z}} |0_{i}\rangle \kr{groundstatefunction}
and excited  state electronic wavefunctions $|e_i\rangle$ with a wave functions of displaced harmonic oscillators
\zr |e_i\Pi_{z}\tilde{n}_{i,z}\rangle= |e_i\rangle \otimes \prod_{z}\frac{1}{\sqrt{n_{i,z}!}}\left(\hat{V}_{i,z}^{\dag}-\sqrt{\frac{m_{i,z}\omega_{i,z}}{2\hbar}}d_{i,z}\right)^{n_{i,z}} |\tilde{0}_{i}\rangle \kr{excitedstatefunction}
where  $|0_{i}\rangle$ is the vibrational ground state on electronic ground state and $|\tilde{0}_{i}\rangle$ is the (shifted) vibrational ground state on electronic excited state. In absence of coupling $J_{ij} = 0$ Hamiltonian (\ref{vibron_ham}) is diagonal in the product basis
$\otimes_i|\psi_i\rangle$ where $|\psi_i\rangle$ is some vector listed in Eq.(\ref{groundstatefunction}) or Eq. (\ref{excitedstatefunction})
$|\psi_i\rangle \in \{ |g_i \Pi_{z} n_{i,z}\rangle , |e_i\Pi_{z}\tilde{n}_{i,z}\rangle \}$.

In a general situation $J_{ij}\neq 0$ molecular Hamiltonian (Eq. (\ref{vibron_ham})) is no longer diagonal in this basis, however the basis still forms a convenient starting point for numerical implementations. Indeed, the matrix element  for coupling terms are composed of the widely-known Franck-Condon factors $\langle n'_{iz}|\tilde{n}_{iz} \rangle$ and can be readily implemented along with diagonal part of the rest of Hamiltonian, and subjected to a standard routines for numerical diagonalization.

The diagonalization of the full vibronic  Hamiltonian $\hat{H}_S $ must be thus,  in general, diagonalized numerically.
\zr \hat{H}_S = \sum_{\alpha} \hbar \varepsilon_{\alpha} |\alpha\rangle \langle \alpha| \kr{bas}
Eigenstates $|\alpha\rangle$ and eigenfrequencies $\varepsilon_{\alpha}$ will be hereafter indexed by Greek letters. Hamiltonian Eq. (\ref{vibron_ham}) conserves number of electronic excitations. Each exciton manifold can be thus diagonalized separately.

On the electronic ground state manifold, there is no resonant coupling, thus the eigenfunctions are direct product $|\Pi_{i,z} g_i n_{i,z} \rangle$ of  ground state electronic wave function with a  well-known wave functions of harmonic oscillators introduced in Eq (\ref{groundstatefunction}). Corresponding eigenenergies are $\hbar \varepsilon_{\gamma} = \hbar \sum_{i,z} n_{i,z} \omega_{i,z}$.  The diagonalization of the single excited electronic manifold is more complex, eigenstates are always obtained by a numeric diagonalization of the corresponding block of the Hamiltonian.
For double excited electronic manifold the situation is different for dimer and longer aggregates.
In a special case of dimer,  doubly excited states does not allow transport (and resonance coupling) and eigenstates are direct product $|e_1 e_2 \Pi_z \tilde{n}_{1,z} \tilde{n}_{2,z} \rangle$ of doubly excited electronic wave function with a wave functions of displaced harmonic oscillators with eigenenergies $\varepsilon_{\delta} = \hbar\epsilon_1 + \hbar\epsilon_2 + \hbar\sum_z (n_{1,z}\omega_{1,z} + n_{2,z}\omega_{2,z})$. For larger aggregates the diagonalization of the second manifold is again numerical. Third and higher manifolds, which may appear in larger aggregates, do not enter calculation of the third order response \cite{abramavicius2009coherent}.

Interaction with the probing laser fields $\vec{E}(t)$ will be treated in a dipole approximation and in Condon approximation described using interaction Hamiltonian
\zr \hat{H}_I(t) =  - \hat{\vec{\mu}} \cdot \vec{E}(t)  \kr{HI}
where $\hat{\vec{\mu}} =  \sum_i \vec{\mu}_i (\hat{A}^{\dag}_i + \hat{A}_i)$; and $\vec{\mu}_i$ is a transition dipole moment between the ground and the excited state of $i$-th molecule.
Its matrix elements
\[
\vec{\mu}_{\alpha \beta}= \sum_i \vec{\mu}_i \langle \alpha |\hat{A}_i |\beta \rangle + c.c.
\]
will be used hereafter. For the section \ref{interactionwithbath}-\ref{liouvillespacepathways} we neglect the vector structure and consider dipole moment as a real number. This approach is sufficient for describing the optical response from most common aggregates with (anti-) parallel dipoles. The neglect will be cured in section \ref{rotationalaveraging}, where the tensorial structure will be calculated together with rotational averaging of the response.

\section{Interaction with bath} \label{interactionwithbath}

The environmental fluctuations modulating system Hamiltonian are mainly of solvent origin or come from less important vibrations. These are modelled by  a dense set of bath harmonic oscillators with bath Hamiltonian
  \zr \hat{H}_B = \sum_k \hbar \Omega_k \hat{B}_k^{\dag} \hat{B}_k\kr{HB}
    where $\Omega_k$ is a frequency of a $k$-bath mode, $\hat{B}_k^{\dag}$ and $ \hat{B}_k$ are standard bosonic creation and annihilation operator, respectively.

System-bath interaction is responsible for damping of vibrational modes and electronic dephasing. Three different couplings are included into $\hat{H}_{SB}$
 \zr \hat{H}_{SB} = \hat{H}_{SB,LD}+\hat{H}_{SB,N} +\hat{H}_{SB,ED} \kr{OOO}
 where the three terms are responsible for Landau-Teller vibrational relaxation, non-linear vibronic-bath couplings, and for electronic dephasing, respectively.
\begin{align}
\hat{H}_{SB,LD}&=\hbar \sum_{k,i,z}  \Omega_k \kappa_{k,i,z}  \left( \hat{B}_k \hat{V}_{i,z}^{\dag}  + \hat{B}_k^{\dag} \hat{V}_{i,z}  \right)  \nonumber\\
\hat{H}_{SB,N} &=  \hbar \sum_{k,i,z}  \Omega_k \zeta_{k,i,z} \left( \hat{B}_k \hat{V}^{\dag}_{i,z}\hat{V}_{i,z} + \hat{B}_k^{\dag} \hat{V}^{\dag}_{i,z}\hat{V}_{i,z} \right) \nonumber\\
\hat{H}_{SB,ED} &= \hbar \sum_{k,i} \Omega_k \iota_{k,i} \left( \hat{B}_k + \hat{B}_k^{\dag}  \right) \hat{A}^{\dag}_i \hat{A}_i
\end{align}
Note that $\hat{H}_{SB,ED}$ and $\hat{H}_{SB,LD}$ are constant and linear, respectively in vibrational coordinate and are thus within the standard spin-boson model.
Such a dynamics can thus be directly compared to standard simulations, they use the same Hamiltonian, but define system and bath differently.
In contrast, the nonlinear coupling $\hat{H}_{SB,N}$ takes us beyond the standard spin-boson model, and there is no correspondence to standard simulations with all vibrations included in bath.

Following the strategy of \cite{mukamel2004many} system-bath Hamiltonian $\hat{H}_{SB}$ shall be divided
$$\hat{H}_{SB}=\hat{H}^D_{SB}+\hat{H}^{OD}_{SB}$$
into the diagonal $\hat{H}^D_{SB}$ and off-diagonal $\hat{H}^{OD}_{SB}$ part of fluctuations responsible for lineshapes and exciton transport respectively. To that end we shall transform system-bath Hamiltonian Eq. (\ref{OOO}) into eigenbasis (Eq. (\ref{bas})) and isolate the diagonal part representing  eigenenergy fluctuations to yield
\zr \hat{H}^D_{SB} =  \hbar \sum_{\alpha} \sum_{k,i} \Omega_k \left( \hat{B}_k (\sum_z v_{\alpha \alpha}^{i,z} \kappa_{k,i,z} + \sum_z w_{\alpha \alpha}^{i,z}\zeta_{k,i,z} + a_{\alpha \alpha}^i\iota_{k,i}) + h.c. \right)|\alpha\rangle \langle \alpha| \kr{diagonal}

Similarly the off-diagonal system-bath Hamiltonian representing eigenstate fluctuations reads
\zr \hat{H}^{OD}_{SB} =  \hbar \sum_{\alpha}\sum_{\beta\neq \alpha} \sum_{k,i} \Omega_k\left( \hat{B}_k (\sum_z v_{\alpha\beta}^{i,z} \kappa_{k,i,z} + \sum_z w_{\alpha \beta}^{i,z}\zeta_{k,i,z} + a_{\alpha \beta}^i\iota_{k,i}) + h.c. )\right)|\alpha\rangle \langle \beta|\kr{offdiagonal}
where we defined the matrix elements
 $$v^{i,z}_{\alpha \beta} = \langle \alpha | \hat{V}_{i,z}^{\dag}|\beta \rangle$$
 $$w^{i,z}_{\alpha \beta} = \langle \alpha |\hat{V}_{i,z}^{\dag} \hat{V}_{i,z}|\beta \rangle$$
 $$  a^{i}_{\alpha \beta} = \langle \alpha| \hat{A}^{\dag}_{i}\hat{A}_{i}|\beta \rangle . $$

The third-order nonlinear optical response for a system with only diagonal system-bath Hamiltonian could be obtained by using a second cumulant expression. The off-diagonal fluctuations shall be treated by means of a master equation. We followed the strategy that combines the two. Before we give the formula to be implemented we first prepare the relevant master equation.

\section{Eigenstate fluctuations: master equations} \label{eigenstatefluctuations}

The off-diagonal fluctuations will be accounted for by using time convolution-less master equation
\zr \frac{d}{dt}\rho_{\nu \mu} = -i (\varepsilon_{\nu}-\varepsilon_{\mu}) \rho_{\nu \mu}  - \sum_{\delta \beta}R_{\nu\mu,\delta\beta}\rho_{\delta\beta}  \,.\kr{Master}
where $\rho_{\nu \mu}\equiv {\rm Tr}\{ |\mu\rangle\langle \nu| \hat{\rho}\}$ is reduced (system) density matrix and $R$ stands for a relaxation tensor. The relaxation tensor shall be evaluated to the second order in system-bath coupling.
As argued by Redfield \cite{redfield1957theory} only certain terms called secular and obeying \zr \varepsilon_{\nu}-\varepsilon_{\mu}-\varepsilon_{\delta}+\varepsilon_{\beta} =0\kr{secular} contribute significantly.
For aggregates, the eigenfrequencies are not degenerated or systematically built, so that Eq. (\ref{secular}) can only be obeyed when either $\varepsilon_{\mu}=\varepsilon_{\nu}$
and $\varepsilon_{\delta}=\varepsilon_{\beta}$ which term describes population transport, or $\varepsilon_{\nu}=\varepsilon_{\delta}$ and $\varepsilon_{\mu}=\varepsilon_{\beta}$ which term describe coherence decay. We adopt this secular approximation on relaxation tensor $R$ in (\ref{Master}) leaving only terms representing population transfer $(\nu = \mu, \delta = \beta)$ or coherence dephasing $(\nu \neq \mu,\, \nu = \delta \,\, \mu = \beta)$ and neglect coherence transfers, coherence to transfer terms etc. In addition to Redfield argumentation based on relevance, we note that such a choice significantly reduces number of Feynman diagrams involved in calculation of coming sections.

The relaxation tensor is calculated in Tokuyama-Mori formalism \cite{tokuyama1976statistical}. We choose form of  Ref \cite{vcapek2001violation,vcapek1994interplay} in asymptotic limit  and evaluate it  to the second order in $\hat{H}_{SB}$.
\zr R_{\nu \mu, \delta \beta} = K_{\nu \mu \delta \beta} \, {\rm Tr}_B\{|\delta\rangle \hat{\rho}_B \langle \beta| \int_0^{\infty} \rm{d} \tau \,e^{-i \breve{L}_0 \tau} \breve{P}\breve{L} e^{i \breve{Q}\breve{L}_0 \tau} \breve{Q}\breve{L}(|\mu \rangle \langle \nu|)  \} \,,\kr{vychozi}
where $K_{\mu \nu \delta \beta} = (1-\delta_{\nu \delta})\delta_{\nu \mu} \delta_{\delta \beta} + \delta_{\nu \delta} \delta_{\mu \beta}$ represent secular approximation and
where the following superoperator notation is introduced:
\zr \breve{L}\cdot = \frac{1}{\hbar}[\hat{H}_S+\hat{H}_{SB}^{OD}+\hat{H}_B,\cdot]\kr{L}
\zr \breve{L}_0\cdot = \frac{1}{\hbar}[\hat{H}_S+\hat{H}_B,\cdot]\kr{L0}
\zr \breve{P}\cdot = \sum_{|\alpha\rangle\langle\gamma|} |\alpha\rangle\langle \gamma| {\rm Tr}\{|\gamma\rangle \hat{\rho}_B \langle \alpha| \cdot \}\kr{P}
\zr \breve{Q} = 1 - \breve{P} \kr{Q}
where
\zr \hat{\rho}_B = \frac{e^{-\beta \hat{H}_B}}{{\rm Tr}_B\{e^{-\beta \hat{H}_B}\}}\kr{rho_B} is canonical bath density matrix.

Evaluating (\ref{vychozi}) yields :
\begin{align}
&R_{\nu \nu, \delta \delta}= -2\sum_{i}\biggl((1+n(\varepsilon_{\delta}-\varepsilon_{\nu}))\bigl(\sum_z (v_{\delta\nu }^{i,z})^2 \mathcal{V}_{i,z}(\varepsilon_{\delta}-\varepsilon_{\nu}) + \sum_z w_{\nu \delta}^{i,z} w_{\delta\nu}^{i,z} \mathcal{W}_{i,z}(\varepsilon_{\delta}-\varepsilon_{\nu}) +  a_{\nu \delta}^{i}a_{\delta\nu}^{i} \mathcal{A}_{i}(\varepsilon_{\delta}-\varepsilon_{\nu})\bigr)\nonumber\\
&+n(\varepsilon_{\nu}-\varepsilon_{\delta}) \bigl(\sum_z (v_{\nu \delta}^{i,z})^2 \mathcal{V}_{i,z}(\varepsilon_{\nu}-\varepsilon_{\delta}) + \sum_z w_{\nu \delta}^{i,z} w_{\delta \nu}^{i,z} \mathcal{W}_{i,z}(\varepsilon_{\nu}-\varepsilon_{\delta}) + a_{\nu \delta}^{i} a_{\delta \nu}^{i} \mathcal{A}_{i}(\varepsilon_{\nu}-\varepsilon_{\delta}) \bigr)\biggr)
\label{redfpop}
\end{align}
$R_{\nu \nu, \delta \delta}$ is relaxation tensor element responsible for populations intra-manifold dynamics between states $\nu$ and $\delta$. When $\nu = \delta$ we have $R_{\nu \nu, \nu \nu} = -\sum_{\delta \neq \nu} R_{\nu \nu, \delta \delta}$. Element $R_{\nu\mu, \nu \mu}$ represents decoherence

\begin{align}
&R_{\nu \mu, \nu \mu}=\sum_{\gamma \neq \mu} \sum_{i}\biggl(n(\varepsilon_{\gamma}-\varepsilon_{\mu}) \bigl(\sum_z(v_{\gamma \mu}^{i,z})^2 \mathcal{V}_{i,z}(\varepsilon_{\gamma}-\varepsilon_{\mu}) + \sum_zw_{\gamma \mu}^{i,z} w_{\mu \gamma}^{i,z} \mathcal{W}_{i,z}(\varepsilon_{\gamma}-\varepsilon_{\mu}) +  a_{\gamma \mu}^{i} a_{\mu \gamma}^{i} \mathcal{A}_{i}(\varepsilon_{\gamma}-\varepsilon_{\mu}) \bigr) \nonumber\\
&+ (1+ n(\varepsilon_{\mu}-\varepsilon_{\gamma}))\bigl(\sum_z(v_{\mu \gamma}^{i,z})^2 \mathcal{V}_{i,z}(\varepsilon_{\mu}-\varepsilon_{\gamma}) + \sum_z w_{\gamma \mu}^{i,z} w_{\mu \gamma}^{i,z} \mathcal{W}_{i,z}(\varepsilon_{\mu}-\varepsilon_{\gamma}) + a_{\gamma \mu}^{i} a_{\mu \gamma}^{i} \mathcal{A}_{i}(\varepsilon_{\mu}-\varepsilon_{\gamma})\bigr)\biggr) \nonumber\\
&+\sum_{\alpha \neq \nu} \sum_{i} \biggl(n(\varepsilon_{\alpha}-\varepsilon_{\nu})\bigl(\sum_z(v_{\alpha \nu}^{i,z})^2 \mathcal{V}_{i,z}(\varepsilon_{\alpha}-\varepsilon_{\nu}) + \sum_z w_{\nu \alpha}^{i,z} w_{\alpha \nu}^{i,z} \mathcal{W}_{i,z}(\varepsilon_{\alpha}-\varepsilon_{\nu}) + a_{\nu \alpha}^{i} a_{\alpha \nu}^{i} \mathcal{A}_{i}(\varepsilon_{\alpha}-\varepsilon_{\nu})\bigr)\nonumber\\
&+(1+n(\varepsilon_{\nu}-\varepsilon_{\alpha}))\bigl(\sum_z(v_{\nu \alpha}^{i,z})^2 \mathcal{V}_{i,z}(\varepsilon_{\nu}-\varepsilon_{\alpha}) + \sum_z w_{\nu \alpha}^{i,z} w_{\alpha \nu}^{i,z} \mathcal{W}_{i,z}(\varepsilon_{\nu}-\varepsilon_{\alpha}) + a_{\nu \alpha}^{i} a_{\alpha \nu}^{i} \mathcal{A}_{i}(\varepsilon_{\nu}-\varepsilon_{\alpha})\bigr)\biggr)
\label{redfcoh}
\end{align}
where $n(x)$ is the Bose-Einstein distribution
\zr n(x) = \frac{1}{e^{\hbar \beta_T x}-1}  \kr{}
and where we defined spectral densities
\zr \mathcal{V}_{i,z}(x) \equiv  \sum_k \Omega_k^2 \kappa_{k,i,z} \kappa_{k,i,z} \delta(x - \Omega_k) \kr{sd1}
\zr \mathcal{W}_{i,z}(x) \equiv  \sum_k \Omega_k^2 \zeta_{k,i,z} \zeta_{k,i,z} \delta(x - \Omega_k)\kr{sd2}
\zr \mathcal{A}_{i}(x) \equiv  \sum_k \Omega_k^2 \iota_{k,i} \iota_{k,i} \delta(x - \Omega_k)\kr{sd3}
In Eqs (\ref{redfpop}) and (\ref{redfcoh}) we have standardly neglected cross terms $\propto  \kappa\iota,  \iota \zeta,  \kappa_ \zeta $. Spectral densities Eq. (\ref{sd1})-(\ref{sd2}) can be arbitrary positive functions defined on positive semi-axis.  Successful estimation of spectral densities from microscopic foundations, i.e. molecular dynamic simulations are rare \cite{olbrich2010time}. In practice there is handful of popular forms for  spectral densities which are parameterized by fitting them to spectra or transfer rates.

For our simulations  we used spectral densities of overdamped Brownian oscillator (exponential decay)
\zr \mathcal{V}_{i,z}(x)  = \frac{2\lambda^{i,z}_{\mathcal{V}} \Lambda^{i,z}_{\mathcal{V}} x}{x^2 + (\Lambda^{i,z}_{\mathcal{V}})^2}\Theta(x)\kr{}
\zr \mathcal{W}_{i,z}(x)  = \frac{2\lambda^{i,z}_{\mathcal{W}} \Lambda^{i,z}_{\mathcal{W}} x}{x^2 + (\Lambda^{i,z}_{\mathcal{W}})^2}\Theta(x)\kr{}
\zr \mathcal{A}_{i}(x)  = \frac{2\lambda^{i}_{\mathcal{A}} \Lambda^{i}_{\mathcal{A}} x}{x^2 + (\Lambda^{i}_{\mathcal{A}})^2}\Theta(x)\kr{}
where $\Lambda_{\mathcal{A(V,W)}}^{i,z}$ is a relaxation rate and $\lambda_{\mathcal{A(V,W)}}^{i,z}$ is a reorganization energy. $\Theta(x)$ denotes Heaviside's step function.

\section{ Eigenfrequency fluctuations: Lineshape functions} \label{eigenfrequencyfluctuations}
We next inquire the effect of diagonal part of system-bath interaction Hamiltonian.
These are responsible for eigenfrequency $\varepsilon_{\alpha}$ represented by bath-space operator
\zr \Delta_{\alpha} \equiv \frac{1}{\hbar} \langle\alpha| \hat{H}_{SB}^D |\alpha \rangle=  \sum_{k,i} \Omega_k \hat{B}_k\left(\sum_z v_{\alpha\alpha}^{i,z} \kappa_{k,i,z}  + \sum_z w_{\alpha\alpha}^{i,z} \zeta_{k,i,z} + a_{\alpha\alpha}^{i}\iota_{k,i}\right) + h.c.\kr{fluctuating_gap}

Our $\hat{B}$-linear coupling introduces Gaussian fluctuations, which can be
fully characterized in terms of  matrix of correlation function
\zr
C_{\beta\alpha}(t) \equiv \left\langle e^{i/\hbar \hat{H}_B t}  \Delta_{\beta} e^{-i/\hbar \hat{H}_B t}  \Delta_{\alpha} \right\rangle_B
\kr{corrf}
where $\langle \hat{X} \rangle_B \equiv Tr_B \{ \hat{X}\rho_B \}$ with $\rho_B$ define in (\ref{rho_B}).
Inserting (\ref{fluctuating_gap}) into (\ref{corrf}) yields
\begin{align}
&C_{\beta \alpha} (t) =  \sum_{k,i} \Omega_k^2 \bigl(\sum_zv_{\beta\beta}^{i,z} \kappa_{k,i,z} + a_{\beta\beta}^{i} \iota_{k,i} + \sum_zw_{\beta\beta}^{i,z} \zeta_{k,i,z} \bigr)\bigl(\sum_zv_{\alpha\alpha}^{i,z}\kappa_{k,i} + a_{\alpha\alpha}^{i}\iota_{k,i} +\sum_z w_{\alpha\alpha}^{i,z} \zeta_{k,i,z}\bigr)\times \nonumber \\
&\times\bigl(\cos{(\Omega_k t)}\coth\left(\frac{\hbar \Omega_k}{2k_B T}\right) -i\sin(\Omega_k t)\bigr) \label{korelacni_funkce_t}
\end{align}

We can calculate (\ref{korelacni_funkce_t}) in terms of spectral densities
\zr C_{\beta \alpha}(t) = \frac{1}{2\pi} \int_{-\infty}^{\infty} d\omega \cos(\omega t) \coth \left(\frac{ \hbar \omega}{2k_BT}\right) C_{\beta\alpha}''(\omega) - \frac{i}{2\pi}\int_{-\infty}^{\infty} \sin(\omega t) C_{\beta \alpha}''(\omega) \kr{}

where
\begin{align}
C_{\beta\alpha}''(\omega)&=\sum_{z,i}v_{\alpha\alpha}^{i,z} v_{\beta\beta}^{i,z}\left[\mathcal{V}_{i,z}(\omega)-\mathcal{V}_{i,z}(-\omega)\right] +\nonumber\\
 &+\sum_{z,i} w_{\alpha\alpha}^{i,z} w_{\beta\beta}^{i,z}\left[\mathcal{W}_{i,z}(\omega)-\mathcal{W}_{i,z}(-\omega)\right] + \sum_i a_{\alpha\alpha}^i a_{\beta\beta}^i\left[\mathcal{A}_{i,z}(\omega)-\mathcal{A}_{i,z}(-\omega)\right]
\end{align}

The effect of Gaussian noise on lineshapes is infamously given by line-broadening function
 $g_{\beta\alpha}(t)$ which can be obtained by a double integration of $C_{\beta\alpha}(t)$
\begin{align}
&g_{\beta\alpha}(t) =  \int_0^t d\tau \int_0^{\tau} d \tau' C_{\beta\alpha}(\tau') = \nonumber\\
&= \frac{1}{2\pi} \int_{-\infty}^{\infty} d\omega \frac{1-\cos{\omega t}}{\omega^2} \coth\left(\frac{\hbar \omega}{2k_B T}\right) C_{\beta\alpha}''(\omega) -\frac{i}{2\pi} \int_{-\infty}^{\infty} d\omega \frac{\sin{\omega t}-\omega t}{\omega^2} C_{\beta\alpha}''(\omega) \label{lineshape}
\end{align}
Now we recall our spectral functions $\mathcal{V}_{i,z}(\omega)$, $\mathcal{W}_{i,z}(\omega)$, $\mathcal{A}_{i,z}(\omega)$ posses  form of a Brownian spectral density
and we obtain well-known result for overdamped Brownian oscillator. For $t>0$:
\begin{align}
g_{\beta\alpha}(t) = & g_{\beta\alpha}'(t) + ig_{\beta\alpha}''(t)\\
g_{\beta\alpha}''(t) =& \sum_{i,z} -(v_{\alpha\alpha}^{i,z} v_{\beta\beta}^{i,z})\lambda_{\mathcal{V}}^{i,z}/\Lambda_{\mathcal{V}}^{i,z} (e^{-\Lambda_{\mathcal{V}}^{i,z} t} + \Lambda_{\mathcal{V}}^{i,z} t -1) -\nonumber\\
&-\sum_i(a_{\alpha\alpha}^i a_{\beta\beta}^i)\lambda_{\mathcal{A}}/\Lambda_{\mathcal{A}}^{i} (e^{-\Lambda_{\mathcal{A}}^{i} t} + \Lambda_{\mathcal{A}}^{i} t -1) -\nonumber\\
&-\sum_{i,z}(w_{\alpha\alpha}^{i,z} w_{\beta\beta}^{i,z})\lambda_{\mathcal{W}}^{i,z}/\Lambda_{\mathcal{W}}^{i,z} (e^{-\Lambda_{\mathcal{W}}^{i,z} t} + \Lambda_{\mathcal{W}}^{i,z} t -1)\\
g'_{\beta\alpha}(t) = &\sum_{i,z} (v_{\alpha\alpha}^{i,z} v_{\beta\beta}^{i,z})\lambda_{\mathcal{V}}^{i,z} /\Lambda_{\mathcal{V}}^{i,z} \coth\left(\frac{\hbar \Lambda_{\mathcal{V}}^{i,z}}{2k_B T}\right)(e^{-\Lambda_{\mathcal{V}}^{i,z} t} + \Lambda_{\mathcal{V}}^{i,z} t - 1) +\nonumber\\
&+\sum_i(a_{\alpha\alpha}^i a_{\beta\beta}^i)\lambda_{\mathcal{A}}^{i}/\Lambda_{\mathcal{A}}^{i} \coth\left(\frac{\hbar \Lambda_{\mathcal{A}}^{i}}{2k_B T} \right)(e^{-\Lambda_{\mathcal{A}}^{i} t} + \Lambda_{\mathcal{A}}^{i} t - 1) + \nonumber\\
&+\sum_{i,z}(w_{\alpha\alpha}^{i,z} w_{\beta\beta}^{i,z})\lambda_{\mathcal{W}}^{i,z} /\Lambda_{\mathcal{W}}^{i,z} \coth\left(\frac{\hbar \Lambda_{\mathcal{W}}^{i,z}}{2k_B T} \right)(e^{-\Lambda_{\mathcal{W}}^{i,z} t} + \Lambda_{\mathcal{W}}^{i,z} t - 1) + \nonumber\\
+& \sum_{i}\left(\sum_z(v_{\alpha\alpha}^{i,z} v_{\beta\beta}^{i,z} + w_{\alpha\alpha}^{i,z} w_{\beta\beta}^{i,z}) + a_{\alpha\alpha}^i a_{\beta\beta}^i\right)\frac{4 \pi \lambda \Lambda k_B T}{\hbar} \sum_{n=1}^{\infty} \frac{e^{-\nu_n t} + \nu_n t -1}{\nu_n(\nu_n^2 - \Lambda^2)} \label{gfce}
\end{align}
where
$\nu_n = \frac{2 \pi}{\hbar \beta} n$ are Matsubara frequencies. To get cumulant at negative times, we use the relation $g_{\beta\alpha}(-t)=g^*_{\beta\alpha}(t)$.

\section{Liouville space pathways} \label{liouvillespacepathways}
Description of exciton population transport by using master equation requires to switch formally into the  Liouville space \cite{mukamel1999principles}. To that end
we introduce superoperators (L) operating on ket (e.g. $\breve{\mu}^{(L)}\hat{X}= \hat{\mu} \hat{X}$) and (R) operating on bra ($\breve{\mu}^{(R)}\hat{X}= \hat{X}\hat{\mu} $)
index of density matrix. Expanding commutator as
$[\hat{\mu},\ldots]= \breve{\mu}^{(L)}-\breve{\mu}^{(R)}$  the linear response function reads
\zr S_L(t) = Tr\{\hat{\mu} \breve{G}(t) \breve{\mu}^{(L)} \hat{\rho}_{eq}\} \kr{linresp}
and six contributions  for photon echo and free induction decay signals of third order response are
\begin{align}
S_1(t_1,t_2,t_3) &= Tr\{ \hat{\mu} \breve{G}(t_3) \breve{\mu}^{(L)} \breve{G}(t_2) \breve{\mu}^{(R)} \breve{G}(t_1) \breve{\mu}^{(R)} \hat{\rho}_{eq} \} \nonumber\\
S_2(t_1,t_2,t_3) &= Tr\{ \hat{\mu} \breve{G}(t_3)  \breve{\mu}^{(R)}\breve{G}(t_2) \breve{\mu}^{(L)} \breve{G}(t_1) \breve{\mu}^{(R)} \hat{\rho}_{eq}  \} \nonumber\\
S_3(t_1,t_2,t_3) &=  Tr\{ \hat{\mu} \breve{G}(t_3) \breve{\mu}^{(L)}\breve{G}(t_2) \breve{\mu}^{(L)} \breve{G}(t_1) \breve{\mu}^{(L)} \hat{\rho}_{eq}  \} \nonumber\\
S_4(t_1,t_2,t_3) &= Tr\{ \hat{\mu} \breve{G}(t_3) \breve{\mu}^{(R)} \breve{G}(t_2) \breve{\mu}^{(R)} \breve{G}(t_1) \breve{\mu}^{(L)} \hat{\rho}_{eq}  \}\nonumber\\
S_5(t_1,t_2,t_3) &= Tr \{ \hat{\mu} \breve{G}(t_3) \breve{\mu}^{(L)}\breve{G}(t_2) \breve{\mu}^{(R)} \breve{G}(t_1) \breve{\mu}^{(L)} \hat{\rho}_{eq}  \} \nonumber\\
S_6(t_1,t_2,t_3) &= Tr\{ \hat{\mu} \breve{G}(t_3) \breve{\mu}^{(L)} \breve{G}(t_2) \breve{\mu}^{(L)} \breve{G}(t_1) \breve{\mu}^{(R)}  \hat{\rho}_{eq}  \}\label{3rdresponse_formal}
\end{align}
where $\breve{G}(t)= e^{-i/\hbar \breve{H}^{(L)} t} e^{-i/\hbar \breve{H}^{(R)} t}$ is the evolution superoperator. Equations (\ref{3rdresponse_formal}) are still quite formal since the evolution superoperator  must include both the master equation and the dynamics of diagonal fluctuations, what is complex task for pathways changing its exciton index in population transfer during a waiting time period. We approximate $G_{\alpha\beta,\gamma\delta}(t)$ as follows

 \begin{equation}
 G_{\alpha\beta,\gamma\delta}(t) = \begin{dcases*}
 \mathcal{G}_{\alpha \beta,\alpha\beta} (t) \mathcal{U}^{L}_{\alpha}(t)  \mathcal{U}^{R*}_{\beta}(t)  & $\alpha = \gamma,\quad \beta = \delta$\\
 \mathcal{G}_{\alpha \alpha,\gamma \gamma} (t) {\mathcal U}^{L}_{\alpha}(t/2)  {\mathcal U}^{R*}_{\alpha}(t/2){\mathcal U}^{L}_{\gamma}(t/2)  {\mathcal U}^{R*}_{\gamma}(t/2) & $\alpha =  \beta,\quad \gamma = \delta, \quad \alpha \neq \gamma$\\
  0 & otherwise
  \end{dcases*}
\label{approx}
  \end{equation}
where ${\mathcal U}_{\alpha}(t)\equiv \langle \alpha| e^{\frac{i}{\hbar}\hat{H}_B t} e^{-\frac{i}{\hbar}(\breve{{H}}_{SB}^D + \breve{{H}}_B) t}|\alpha\rangle$ is evolution of bath in $\alpha$ eigenstate written in interaction picture and $\breve{\mathcal{G}}$ is Green function of master equation (\ref{Master}).

With use of Eq. (\ref{linresp}) linear response function  reads
\zr S_L(t) = \sum_{\alpha\beta} \mu_{\alpha\beta} \mathcal{G}_{\beta\alpha,\beta\alpha}(t)\mu_{\beta\alpha}\rho_{\alpha}F_{\alpha\beta}^2(t,0)\kr{linearresponse}
where $\rho_{eq,\alpha}\equiv {\rm Tr_B}\{\langle\alpha|\hat{\rho}_{eq}|\alpha\rangle\}$ is total density in state $\alpha$ and we merged bath phase factor into
\zr F_{\beta\alpha}^2(\tau_2,\tau_1) \equiv {\rm Tr_B}\left\{ \mathcal{U}_{\alpha}(0-\tau_{2}) \mathcal{U}_{\beta}(\tau_{2}-\tau_{1}) \mathcal{U}_{\alpha}(\tau_1-0)\langle\alpha|\hat{\rho}_{eq}|\alpha\rangle \right\} \kr{lindef}
The equilibrium density matrix $\hat{\rho}_{eq}$ is defined as 
\zr \hat{\rho}_{eq} \equiv  \frac{e^{-\frac{\hat{H}_v+\hat{H}_{SB}+\hat{H}_{B}}{k_BT}}}{{\rm Tr}\left\{e^{-\frac{\hat{H}_v+\hat{H}_{SB}+\hat{H}_B}{k_BT}} \right\}}|\Pi_i g_i\rangle\langle\Pi_i g_i|\kr{rho_eq}
where we assumed that the process starts at chromophore's electronic ground state with vibrational level populated according to Boltzman distribution. Using Gell-Mann's theorem \cite{gell1951bound} we can obtain the equilibrium density matrix by switching the interaction $\hat{H}_{SB}$ at time $t=-\infty$ where we start from uncorrelated density matrix $\frac{e^{-\frac{\hat{H}_v}{k_BT}}}{{\rm Tr}\left\{e^{-\frac{\hat{H}_v}{k_BT}}\right\}}|\Pi_i g_i\rangle\langle\Pi_i g_i| \otimes \hat{\rho}_B$, i.e.
\begin{align}
\hat{\rho}_{eq} &= \mathcal{U}(0-\infty) \frac{e^{-\frac{\hat{H}_v}{k_BT}}}{{\rm Tr}\left\{e^{-\frac{\hat{H}_v}{k_BT}}\right\}}|\Pi_i g_i\rangle\langle\Pi_i g_i| \otimes \hat{\rho}_B \mathcal{U}^*(0-\infty) \approx\nonumber\\
&\approx \sum_{\alpha \in g} \mathcal{U}_{\alpha}(0-\infty) |\alpha\rangle\langle\alpha| \frac{e^{\frac{-\hbar \varepsilon_{\alpha}}{k_BT}}}{\sum_{\alpha \in g} e^{\frac{-\hbar \varepsilon_{\alpha}}{kBT}}}\otimes\hat{\rho}_B \mathcal{U}^*_{\alpha}(0-\infty) \label{rho_app}
\end{align}
where the index $\alpha$ runs over vibrational levels of electronic ground state $g$, $\mathcal{U}(t)\equiv  e^{\frac{i}{\hbar}\hat{H}_B t} e^{-\frac{i}{\hbar}(\breve{{H}}_{SB} + \breve{{H}}_B) t}$ was approximated $\langle\alpha|\mathcal{U}(t)|\alpha\rangle \approx \mathcal{U}_{\alpha}(t)$. That leads us to
\zr F_{\beta\alpha}^2(\tau_2,\tau_1)= \left\langle \mathcal{U}_{\alpha}(-\tau_{2}-\infty) \mathcal{U}_{\beta}(\tau_{2}-\tau_{1}) \mathcal{U}_{\alpha}(\tau_1+\infty) \right\rangle_B \kr{F2}
and 
\zr \rho_{eq,\alpha}= \frac{e^{\frac{-\hbar \varepsilon_{\alpha}}{k_BT}}}{\sum_{\alpha \in g} e^{\frac{-\hbar \varepsilon_{\alpha}}{kBT}}}\kr{} 
Formula (\ref{F2}) for bath phase factor can be evaluated exactly by using the second cumulant \cite{mukamel1983nonimpact}
\begin{align}
F_{\beta\alpha}^2(\tau_2,\tau_1) &= \exp\left(-g_{\alpha\alpha}(\tau_{1}+\infty)-g_{\beta\beta}(\tau_{2}-\tau_{1}) - g_{\alpha\alpha}(-\infty-\tau_{2}) +\right.\nonumber\\
&+ g_{\alpha\beta}(\tau_{1}+\infty) +g_{\alpha\beta}(\tau_{2}-\tau_{1})-g_{\alpha\beta}(\tau_{2}+\infty)-g_{\alpha\alpha}(\tau_{2}-\tau_{1}) + g_{\alpha\alpha}(\tau_{2}+\infty)+\nonumber\\
&\left.+g_{\alpha\alpha}(-\infty-\tau_1) + g_{\beta\alpha}(\tau_{2}-\tau_1) + g_{\beta\alpha}(-\infty -\tau_2) -g_{\beta\alpha}(-\infty -\tau_1) \right)\label{F2complete}
\end{align}

Note, that $\alpha$  always represent the initial state.  
To simplify formula for bath phase factors we note, that the bath evolution in initial state is trivial because in most realistic cases $\Delta_{\alpha}=0$.  For instance it holds for all electronic ground state levels for $\hat{H}_{SB,ED}$ and $\hat{H}_{SB,LD}$ Hamiltonians, and for the lowest vibrational level of $\hat{H}_{SB,N}$ (i.e. at low temperatures). Then
 $\mathcal{U}_{\alpha} =1$ and Eq. (\ref{F2}) can be simplified to form in which we dropped trivial evolution in $\alpha$ state
$$F_{\beta\alpha}^2(\tau_2,\tau_1) \bigg|_{\Delta_{\alpha}=0}= \left\langle  \mathcal{U}_{\beta}(\tau_{21}) \right\rangle_B = \exp(-g_{\beta\beta}(\tau_{21}))$$
where $\tau_{ij}\equiv \tau_i-\tau_j$ and we will keep this level of approximation for higher lineshape functions.

Accumulating bath phase factors for evolution along exciton transfer pathways $G_{\alpha \alpha,\gamma\gamma}$
according Eq. (\ref{approx}) is approximate. We assumed that bath is evolving in the $\gamma$ state for the first half of interval $t_2$, at which point it jumps to $\alpha$ state and evolve here for the rest of interval.

\begin{figure}[htb]
  \begin{center}
    \includegraphics[width=\textwidth]{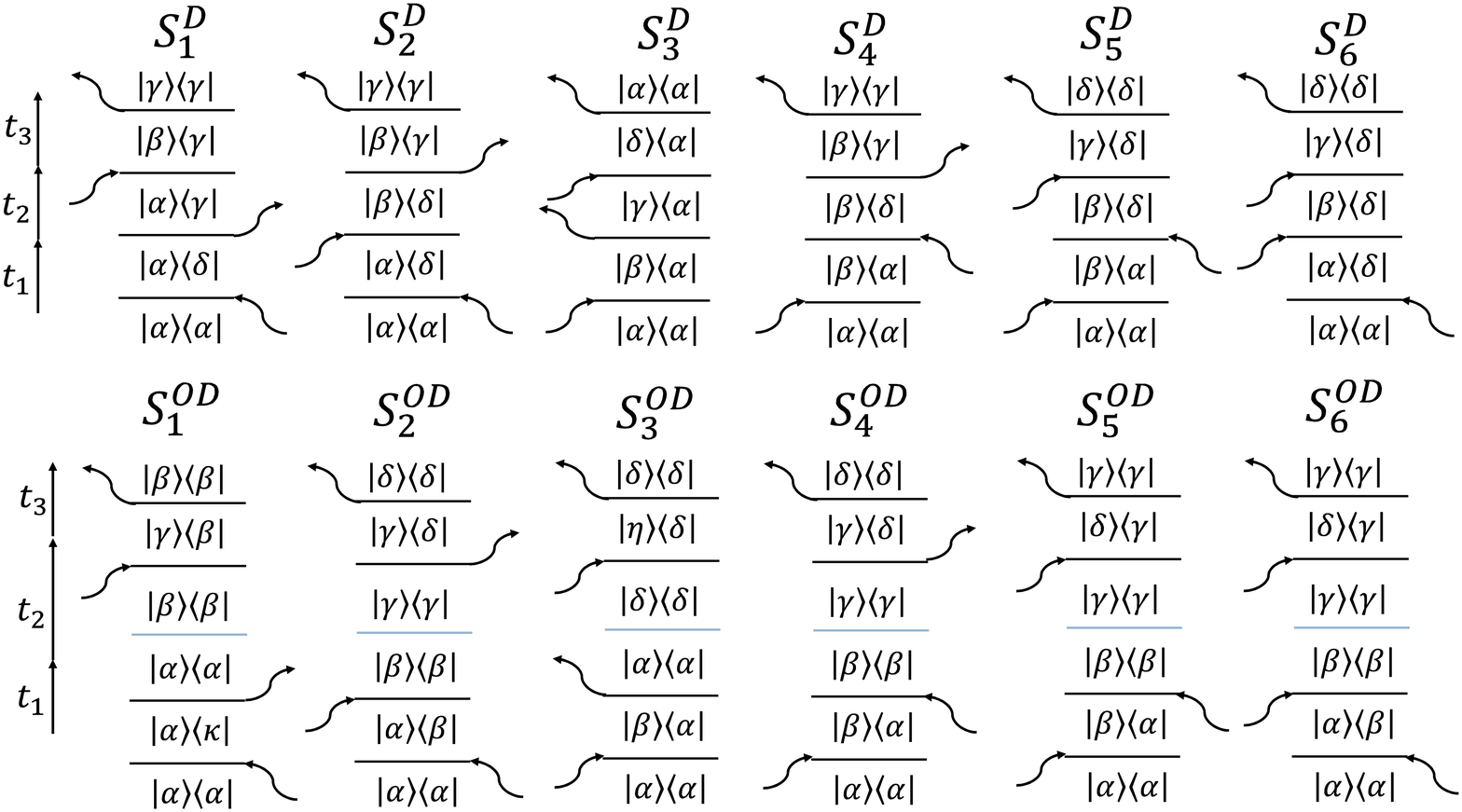}
  \caption{Feynman diagrams for Liouville space pathways contributing to the third-order response signal photon echo $k_I$ and $k_{II}$ of a vibronic system under assumptions of the rotating wave approximation. Indices $\alpha,\beta,\delta,\gamma$ index eigenstates.}
    \label{Feynman}
  \end{center}
\end{figure}

As the transfers can only occur during the second (waiting) time interval $t_2$,  we partition response Eq (\ref{3rdresponse_formal}) to contributions from  transfer $S_{OD}$ and non-transfer $S_D$  pathways as visualized by 12 Feynman diagrams on Fig. \ref{Feynman}

\zr S_j(t_1,t_2,t_3) =  S^D_j(t_1,t_2,t_3) + S^{OD}_j(t_1,t_2,t_3)\kr{}

Expanding Eq. (\ref{3rdresponse_formal})  in the excitonic index and merging phase factors we get
\begin{align*}
S_1^D(t_1,t_2,t_3) &= \sum_{\alpha,\beta,\gamma,\delta}  \mu_{\gamma\beta} \mathcal{G}_{\beta\gamma,\beta\gamma}(t_3) \mu_{\beta\alpha} {\mathcal G}_{\alpha\gamma,\alpha\gamma}(t_2) \mu_{\delta\gamma} {\mathcal G}_{\alpha\delta,\alpha\delta}(t_1)\mu_{\alpha\delta} \rho_{eq,\alpha} F^{4}_{\delta,\gamma,\beta,\alpha}(0,t_1,t_1+t_2+t_3,t_1+t_2) \\
S_2^D(t_1,t_2,t_3) &= \sum_{\alpha,\beta,\gamma,\delta}  \mu_{\gamma\beta} \mathcal{G}_{\beta\gamma,\beta\gamma}(t_3) \mu_{\delta\gamma} {\mathcal G}_{\beta\delta,\beta\delta}(t_2) \mu_{\beta\alpha} {\mathcal G}_{\alpha\delta,\alpha\delta}(t_1)\mu_{\alpha\delta} \rho_{eq,\alpha}  F^{4}_{\delta,\gamma,\beta,\alpha}(0,t_1+t_2,t_1+t_2+t_3,t_1) \\
S_3^D(t_1,t_2,t_3) &= \sum_{\alpha,\beta,\gamma,\delta}  \mu_{\alpha\delta} \mathcal{G}_{\delta\alpha,\delta\alpha}(t_3) \mu_{\delta\gamma} {\mathcal G}_{\gamma\alpha,\gamma\alpha}(t_2) \mu_{\gamma\beta} {\mathcal G}_{\beta\alpha,\beta\alpha}(t_1)\mu_{\beta\alpha} \rho_{eq,\alpha}  F^{4}_{\alpha,\beta,\gamma,\delta}(t_1+t_2+t_3,t_1+t_2,t_1,0) \\
S_4^D(t_1,t_2,t_3) &= \sum_{\alpha,\beta,\gamma,\delta}  \mu_{\gamma\beta} \mathcal{G}_{\beta\gamma,\beta\gamma}(t_3) \mu_{\delta\gamma} {\mathcal G}_{\beta\delta,\beta\delta}(t_2) \mu_{\alpha\delta} {\mathcal G}_{\beta\alpha,\beta\alpha}(t_1)\mu_{\beta\alpha} \rho_{eq,\alpha}  F^{4}_{\delta,\gamma,\beta,\alpha}(t_1,t_1+t_2,t_1+t_2+t_3,0) \\
S_5^D(t_1,t_2,t_3) &= \sum_{\alpha,\beta,\gamma,\delta}  \mu_{\delta\gamma} \mathcal{G}_{\gamma\delta,\gamma\delta}(t_3) \mu_{\gamma\beta} {\mathcal G}_{\beta\delta,\beta\delta}(t_2) \mu_{\alpha\delta} {\mathcal G}_{\beta\alpha,\beta\alpha}(t_1)\mu_{\beta\alpha} \rho_{eq,\alpha}  F^{4}_{\delta,\gamma,\beta,\alpha}(t_1,t_1+t_2+t_3,t_1+t_2,0) \\
S_6^D(t_1,t_2,t_3) &= \sum_{\alpha,\beta,\gamma,\delta}  \mu_{\gamma\beta} \mathcal{G}_{\gamma\delta,\gamma\delta}(t_3) \mu_{\gamma\beta} {\mathcal G}_{\beta\delta,\beta\delta}(t_2) \mu_{\beta\alpha} {\mathcal G}_{\delta\alpha,\delta\alpha}(t_1)\mu_{\alpha\delta} \rho_{eq,\alpha}  F^{4}_{\delta,\gamma,\beta,\alpha}(0,t_1+t_2+t_3,t_1+t_2,t_1) \\
S_1^{OD}(t_1,t_2,t_3) &=\sum_{\alpha,\beta,\gamma,\kappa} \mu_{\beta\gamma} \mathcal{G}_{\gamma\beta,\gamma\beta}(t_3) \mu_{\gamma\beta} {\mathcal G}_{\beta\beta,\alpha\alpha}(t_2) \mu_{\kappa\alpha} {\mathcal G}_{\alpha\kappa,\alpha\kappa}(t_1)\mu_{\alpha\kappa} \rho_{eq,\alpha}\times\\
&\times F^{6}_{\kappa,\alpha,\beta,\gamma,\beta,\alpha}(0,t_1,t_1+t_2/2,t_1+t_2+t_3,t_1+t_2,t_1+t_2/2)  \\
S_2^{OD}(t_1,t_2,t_3) &=\sum_{\alpha,\beta,\gamma,\delta} \mu_{\delta\gamma} \mathcal{G}_{\gamma\delta,\gamma\delta}(t_3) \mu_{\gamma\delta} {\mathcal G}_{\gamma\gamma,\beta\beta}(t_2) \mu_{\beta\alpha} {\mathcal G}_{\alpha\beta,\alpha\beta}(t_1)\mu_{\alpha\beta} \rho_{eq,\alpha}\times\\
 &\times F^{6}_{\beta,\gamma,\delta,\gamma,\beta,\alpha}(0,t_1+t_2/2,t_1+t_2,t_1+t_2+t_3,t_1+t_2/2,t_1)\\
S_3^{OD}(t_1,t_2,t_3) &=\sum_{\alpha,\beta,\delta,\eta} \mu_{\delta\eta} \mathcal{G}_{\eta\delta,\eta\delta}(t_3) \mu_{\eta\delta} {\mathcal G}_{\delta\delta,\alpha\alpha}(t_2) \mu_{\alpha\beta} {\mathcal G}_{\beta\alpha,\beta\alpha}(t_1)\mu_{\beta\alpha} \rho_{eq,\alpha}\times\\
 &\times F^{6}_{\delta,\eta,\delta,\alpha,\beta,\alpha}(t_1+t_2/2,t_1+t_2+t_3,t_1+t_2,t_1+t_2/2,t_1,0)   \\
S_4^{OD}(t_1,t_2,t_3) &=\sum_{\alpha,\beta,\gamma,\delta} \mu_{\delta\gamma} \mathcal{G}_{\gamma\delta,\gamma\delta}(t_3) \mu_{\gamma\delta} {\mathcal G}_{\gamma\gamma,\beta\beta}(t_2) \mu_{\alpha\beta} {\mathcal G}_{\beta\alpha,\beta\alpha}(t_1)\mu_{\beta\alpha} \rho_{eq,\alpha}\times\\
 &\times F^{6}_{\beta,\gamma,\delta,\gamma,\beta,\alpha}(t_1,t_1+t_2/2,t_1+t_2,t_1+t_2+t_3,t_1+t_2/2,0)   \\
S_5^{OD}(t_1,t_2,t_3) &=\sum_{\alpha,\beta,\gamma,\delta} \mu_{\gamma\delta} \mathcal{G}_{\delta\gamma,\delta\gamma}(t_3) \mu_{\delta\gamma} {\mathcal G}_{\gamma\gamma,\beta\beta}(t_2) \mu_{\alpha\beta} {\mathcal G}_{\beta\alpha,\beta\alpha}(t_1)\mu_{\beta\alpha} \rho_{eq,\alpha}\times\\
 &\times F^{6}_{\beta,\gamma,\delta,\gamma,\beta,\alpha}(t_1,t_1+t_2/2,t_1+t_2+t_3,t_1+t_2,t_1+t_2/2,0)   \\
S_6^{OD}(t_1,t_2,t_3) &=\sum_{\alpha,\beta,\gamma,\delta} \mu_{\gamma\delta} \mathcal{G}_{\delta\gamma,\delta\gamma}(t_3) \mu_{\delta\gamma} {\mathcal G}_{\gamma\gamma,\beta\beta}(t_2) \mu_{\beta\alpha} {\mathcal G}_{\alpha\beta,\alpha\beta}(t_1)\mu_{\alpha\beta} \rho_{eq,\alpha} \times\\
 &\times F^{6}_{\beta,\gamma,\delta,\gamma,\beta,\alpha}(0,t_1+t_2/2,t_1+t_2+t_3,t_1+t_2,t_1+t_2/2,t_1)
\end{align*}

We shall next to evaluate the bath phase factors bath phase factors 
\zr F^4_{\delta,\gamma,\beta,\alpha}(\tau_4,\tau_3,\tau_2,\tau_1) =
\left\langle \mathcal{U}_{\alpha}(-\infty -\tau_4) \mathcal{U}_{\delta}(\tau_{43})\mathcal{U}_{\gamma}(\tau_{32})\mathcal{U}_{\beta}(\tau_{21}) \mathcal{U}_{\alpha}(\tau_1 + \infty)  \right\rangle_B \kr{fundef}
$F^4$ can be again simplified using $\Delta_{\alpha} =0$ to
\zr F^4_{\delta,\gamma,\beta,\alpha}(\tau_4,\tau_3,\tau_2,\tau_1) \bigg|_{\Delta_{\alpha}=0} = \left\langle  \mathcal{U}_{\delta}(\tau_{43})\mathcal{U}_{\gamma}(\tau_{32})\mathcal{U}_{\beta}(\tau_{21}) \right\rangle_B \kr{nonfundef}
For our $B$-linear harmonic bath, Eq (\ref{nonfundef}) can be figured out exactly using second cumulant \cite{mukamel1983nonimpact}
\begin{align}
 &F^4_{\delta,\gamma,\beta,\alpha}(\tau_4,\tau_3,\tau_2,\tau_1) \bigg|_{\Delta_{\alpha}=0}= \exp {\big \{ } -g_{\beta\beta}(\tau_{21})-g_{\gamma\gamma}(\tau_{32})-g_{\delta\delta}(\tau_{43})+g_{\beta\gamma}(\tau_{21})+g_{\beta\gamma}(\tau_{32})-g_{\beta\gamma}(\tau_{31})-\nonumber\\
 &-g_{\beta\delta}(\tau_{32})-g_{\beta\delta}(\tau_{41})+g_{\beta\delta}(\tau_{31})+g_{\beta\delta}(\tau_{42})+g_{\gamma\delta}(\tau_{32})+g_{\gamma\delta}(\tau_{43})-g_{\gamma\delta}(\tau_{42})
 {\big \} }
\end{align}
The full expression (\ref{fundef}) could be evaluated along similar lines, however we did not meet yet realistic need to implement it. For pathways with transfer during $t_2$ interval we proceed in similar manner. We define
\begin{align}
F^6_{\kappa,\eta,\delta,\gamma,\beta,\alpha}(\tau_6,\tau_5,\tau_4,\tau_3,\tau_2,\tau_1) &= \left\langle U_{\alpha}(-\infty-\tau_6)U_{\kappa}(\tau_{65})
U_{\eta}(\tau_{54})U_{\delta}(\tau_{43})U_{\gamma}(\tau_{32}) U_{\beta}(\tau_{21}) U_{\alpha}(\tau_{1}+\infty) \right\rangle_B
\end{align}
which is evaluated as
\begin{align}
&F^6_{\kappa,\eta,\delta,\gamma,\beta,\alpha}(\tau_6,\tau_5,\tau_4,\tau_3,\tau_2,\tau_1) \bigg|_{\Delta_{\alpha}=0} = \exp {\big \{ } -g_{\beta\beta}(\tau_{21})-g_{\gamma\gamma}(\tau_{32})-g_{\delta\delta}(\tau_{43})-g_{\eta\eta}(\tau_{54})-g_{\kappa\kappa}(\tau_{65})-\nonumber\\
&-g_{\beta\gamma}(\tau_{31})+g_{\beta\gamma}(\tau_{32})+g_{\beta\gamma}(\tau_{21})-g_{\beta\delta}(\tau_{41})+g_{\beta\delta}(\tau_{42})+g_{\beta\delta}(\tau_{31})
-g_{\beta\delta}(\tau_{32})-g_{\beta\eta}(\tau_{51})+\nonumber\\
&+g_{\beta\eta}(\tau_{52})+g_{\beta\eta}(\tau_{41})-g_{\beta\eta}(\tau_{42})-g_{\beta\kappa}(\tau_{61})+g_{\beta\kappa}(\tau_{62})+g_{\beta\kappa}(\tau_{51})
-g_{\beta\kappa}(\tau_{52})-g_{\gamma\delta}(\tau_{42})+\nonumber\\
&+g_{\gamma\delta}(\tau_{43})+g_{\gamma\delta}(\tau_{32})-g_{\gamma\eta}(\tau_{52})+g_{\gamma\eta}(\tau_{53})+g_{\gamma\eta}(\tau_{42})-g_{\gamma\eta}(\tau_{43})
-g_{\gamma\kappa}(\tau_{62})+g_{\gamma\kappa}(\tau_{63})+\nonumber\\
&+g_{\gamma\kappa}(\tau_{52})-g_{\gamma\kappa}(\tau_{53})-g_{\delta\eta}(\tau_{53})+g_{\delta\eta}(\tau_{54})+g_{\delta\eta}(\tau_{43})-g_{\delta\kappa}(\tau_{63})
+g_{\delta\kappa}(\tau_{64})+g_{\delta\kappa}(\tau_{53})-\nonumber\\
&-g_{\delta\kappa}(\tau_{54})-g_{\eta\kappa}(\tau_{64})+g_{\eta\kappa}(\tau_{63})+g_{\eta\kappa}(\tau_{54}) {\big \} }
\end{align}
\section{Rotational averaging}\label{rotationalaveraging}

Response functions of section 2-6 are advantageously calculated in molecular frame. In most experiments the molecules are
 randomly oriented in an isotropic sample. The lab frame  predictions require rotationally averaging.
  When the dipoles in aggregate are (anti)parallel the molecular frame response has only single component and the rotational averaging are thus just simple unimportant factors. Normalized spectrum could thus be calculated using the previous sections without any averaging. When the dipole orientation is general, there is more components of the molecular frame tensor, which have to be combined as follows.

The response functions of the $(n-1)$th order are $(n)$-rank tensors and can be averaged according Ref.  \cite{andrews1977three}
\zr {\mathcal S}_{l_1,l_2,\ldots,l_n}^{(n-1)} = I^{(n)}_{l_1,l_2,\ldots,l_n;m_1,m_2,\ldots,m_n}S^{(n-1)}_{m_1,m_2,\ldots,m_n}\kr{}
where $S$ is a molecular frame response, averaged lab frame response ${\mathcal S}$,
 $m_k$ are axes in molecular frame and $I^{(n)}_{l_1,l_2,\ldots,l_n;m_1,m_2,\ldots,m_n}$ is the transformation.

For linear response  $n=2$,
\zr I^{(2)} = \frac{1}{3}\delta_{l_1,l_2}\delta_{m_1,m_2}\kr{}
and the response function relevant to CW absorption is  ${\mathcal S}= S_{xx}+S_{yy}+S_{zz}$.

For the third order response, $n=4$ and
\zr I^{(4)} = \frac{1}{30} \begin{pmatrix} \delta_{l_1,l_2}\delta_{l_3,l_4} \\ \delta_{l_1,l_3}\delta_{l_2,l_4}\\ \delta_{l_1,l_4}\delta_{l_2,l_3} \end{pmatrix}^T \begin{pmatrix} 4 & -1 & -1\\ -1 & 4 & -1\\ -1 & -1& 4 \end{pmatrix}  \begin{pmatrix} \delta_{m_1,m_2}\delta_{m_3,m_4} \\ \delta_{m_1,m_3}\delta_{m_2,m_4}\\ \delta_{m_1,m_4}\delta_{m_2,m_3} \end{pmatrix} \kr{}

In a most common arrangement of the third order experiments all polarizations of the laser fields are in the same direction. In addition, for a dimer we can choose the molecular frame
so that both dipoles are in two directions $x$ and $y$. For this case we need to calculate 8 molecular tensor components $S_{xxxx}$, $S_{yyyy}$, $S_{xxyy}$, $S_{yyxx}$, $S_{xyyx}$, $S_{yxxy}$, $S_{xyxy}$, and $S_{yxyx}$.
The lab frame response function is
\[
{\mathcal S}=\frac{1}{15}(3S_{xxxx}+3S_{yyyy}+S_{xxyy}+S_{yyxx}+S_{xyyx}+S_{yxxy}+S_{xyxy}+S_{yxyx})
\]

\section{Signals} \label{spectracalculations}
We next connect the response functions with experimental quantities.
Weak-field absorption is obtained as a Fourier transform of linear response function $S_L(t)$
\zr S(\Omega)= {\rm Re} \int_0^{\infty}e^{i\Omega t} S_L(t) dt \kr{Absorpce}

 The nonlinear time-domain techniques are calculated in an impulsive limit, i.e. envelope of exciting laser field are approximated by $\delta$-pulse $E(t) = \delta(t)$.
Time resolved fluorescence is emission (represented by diagrams 2,4) delayed by $t_2$ after two interactions with the same exciting pulse  $t_1=0$
\zr S_F(t_2,\Omega)= {\rm Re}  \sum_{n=2,4} \int_0^{\infty}  dt_3 e^{i\Omega t_3 } S_n(t_1=0,t_2,t_3) \kr{fluorescence}

Pump-probe (transient absorption) signal is obtained in FWM experiment when first two interactions are by the same pulse $k_1 = k_2$, $t_1=0$. The spectrum is not resolved  in $\omega_1$  and the signal emerge in the single direction $k_3 = k_S$ combining both  the rephasing and nonrephasing contributions.
\zr S_{PP}(t_2,\Omega)= {\rm Re}\sum_{n=1}^6 \int_0^{\infty}  dt_3 e^{i\Omega t_3 } S_n(t_1=0,t_2,t_3) \kr{pumpprobe}

Transient grating is similar except that the two pump pulses are different $k_1 \neq k_2$, while the time-ordering between the first two pulses is not attained. We can still look into both phase matching directions and manipulate the relative phase of pump pulses, what
adds
the imaginary part of response as an observable
\zr S_{TG}(t_2,\Omega)= \sum_{n=1}^6 \int_0^{\infty}  dt_3 e^{i\Omega t_3 } S_n(t_1=0,t_2,t_3) \kr{TGsignal}
The $t_2$ dynamics of transient grating or pump probe spectra are often mapped onto a generic kinetic schemes and inverse rate constant of certain step is identified as the {\it transfer time} for the jump between donor and acceptor chromophore. This is often used . The exact definition (i.e. specification of the kinetic scheme) of this often used dynamical measure depends on the context. Details of retrieval procedure can be found in \cite{stokkum2004global,lincoln2015A}.

2D spectrograms are results of fully resolved and time-ordered measurements. We define mixed time-frequency representation of response functions \ref{3rdresponse_formal}  as follows
\zr S_n(\Omega_1, t_2, \Omega_3) = \int_0^{\infty} \int_0^{\infty} dt_1 dt_3 e^{i\Omega_3 t_3 + i \Omega_1 t_1} S_n(t_1,t_2,t_3) \kr{spektrogram}

Rephasing signal $k_I$ is given by
\zr S_I(\Omega_1,t_2,\Omega_3) = S_1(\Omega_1,t_2,\Omega_3) + S_2(\Omega_1,t_2,\Omega_3) + S_6(\Omega_1,t_2,\Omega_3) \kr{rephasing}

and for non-rephasing $k_{II}$
\zr S_{II}(\Omega_1,t_2,\Omega_3) = S_3(\Omega_1,t_2,\Omega_3) + S_4(\Omega_1,t_2,\Omega_3) + S_5(\Omega_1,t_2,\Omega_3) \kr{nonrephasing}

We usually display the following combination of signals that provide absorptive signal \cite{khalil2003obtaining}
\zr S_{A}(\Omega_1,t_2,\Omega_3) = {\rm Re} S_I(-\Omega_1,t_2,\Omega_3) + {\rm Re}  S_{II}(\Omega_1,t_2,\Omega_3) \kr{absorptive}
Correlation $(\Omega_1,\Omega_3)$ plots at fixed waiting times $t_2$ typically show a few diagonal $\Omega_1=\Omega_3$ peaks at the frequencies of the vibronic transitions and a few off-diagonal  $\Omega_1 \neq\Omega_3$ peaks, which typically show a harmonic modulation along waiting time $ \propto \sin[(\Omega_1 -\Omega_3) t_2 +\varphi]$. The {\it oscillatory phase} $\varphi$ is another frequently used measure.  Details of phase retrieval from the 2D signal are given in Supplementary Information of Ref. \cite{perlik2014distinguishing}.

\section{Applications}\label{applicatons}
We have used the present code for several recent publications. We have studied relative oscillatory phase between diagonal and cross peaks of 2D spectra and its temperature sensitivity for a generic vibronic dimer (P-model) and monomer (V-model) \cite{perlik2014distinguishing}.
Vibronic dimer was next used as a model for transient grating (TG) studies of exciton transfer from carotenoid's S2 to bacteriochlorophyll Qx \cite{perlik2015vibronic} and similar study for perylene complexes is in preparation \cite{perlik2016perylene}. The code was also used to look into the electronic vibrational dynamics of (monomers of) hypericin \cite{lincoln2015A}
 Details of the application are summarized in the following table.
\begin{center}
\begin{tabularx}{\textwidth}{|c|X|c|c|c|c|c|c|X|}
  \hline
  \multirow{2}{*}{Publ.} & \multirow{2}{*}{System} &\multicolumn{3}{|c|}{Coupling} & \multirow{2}{*}{Aggregate} & \multicolumn{2}{|c|}{Oscillators on}  & \multirow{2}{*}{Signals}\\ \cline{3-5} \cline{7-8}
   & & LD & ED & N & & site 1 & site 2 & \\ \hline
  \multirow{2}{*}{\cite{perlik2014distinguishing}} & V-model &  Y & N & N & monomer & 1 & - &2D, oscillatory phase \\ \cline{2-9}
     & P-model & Y & N & N & dimer & 1 & 1 &2D, oscillatory phase\\ \hline
  \multirow{2}{*}{\cite{perlik2015vibronic}} & LH2 Carotenoid - chlorophyll pair & Y & N & N& dimer & 1 & 0 & transfer time\\ \cline{2-9}
      & Dyad Purpurin - Bchl a& Y & N & N& dimer & 1 & 1 & transfer time\\ \hline
      \cite{perlik2016perylene} & Perylene dyad & Y & Y & Y& dimer & 1 & 1& 2D, TG, transfer time\\ \hline
 \multirow{2}{*}{\cite{lincoln2015A}} & \multirow{2}{*}{Hypericin} & \multirow{2}{*}{Y} & \multirow{2}{*}{Y} & \multirow{2}{*}{N} & \multirow{2}{*}{monomer} & 2 & 0 & absorption \\ \cline{7-9}
       & & & & & & 1 &0 & TG\\ \hline
\end{tabularx}
\end{center}

\section{Conclusion}\label{conclusion}
The present code was primarily designed
 to simulate spectra of molecular dimers with few harmonic oscillators, as summarized in the previous section.
Within these limits the code was thus extensively tested, it works without apparent flaws and it can be routinely used now.
The model outlined in this communication, however, allow for some straightforward extensions. Let us discuss here two of them which could be implemented in near future.

First, extension to a larger aggregates is formally trivial as it can be achieved by extending the limits of the relevant index,
and also the required changes of the code will be similarly simple. The trouble here is the computational cost of such simulations quickly expanding with number of oscillators involved, which will hardly allow extend much beyond trimer (with 3 oscillators).

Next, we have implemented rather general routine for diagonalization of system (exciton-vibration) Hamiltonian, not really bounded to harmonic potential profiles.
In fact, any exciton-vibrational Hamiltonian can be diagonalized, once its matrix elements in the harmonic basis are inserted.
One can thus readily implement more realistic models of vibrational profiles (Morse potential, Lenard-Jones potential etc).

The development of our code would be impossible without advices of our kind collaborators.
We are indebted to Tom\'{a}\v{s} Man\v{c}al, J\"{u}rgen Hauer, Craig Lincoln, and Arpa Galestian Pour, who at various stages  discussed, inspired,  demanded, and otherwise helped to shape our work.

\DeclareRobustCommand{\VAN}[3]{#3}
\printbibliography
\end{document}